\documentclass{article}

\usepackage{epsfig}
\newcommand{\bm}{\bf}

\begin{document}

\begin{center}
{\bf \large Non-spiky density of states of an icosahedral quasicrystal}
\end{center}

\begin{center}
E. S. Zijlstra\thanks{E-mail: eeuwe@sci.kun.nl} and
T. Janssen\thanks{E-mail: ted@sci.kun.nl}
\end{center}

\begin{center}
{\it Institute for Theoretical Physics, University of Nijmegen -\\
Toernooiveld, 6525 ED Nijmegen, The Netherlands}
\end{center}

\begin{abstract}
The density of states of the ideal three-dimensional Penrose tiling, a
quasicrystalline model, is calculated with a resolution of 10~meV. It is not
spiky. This falsifies theoretical predictions so far, that spikes of width
10-20~meV are generic for the density of states of quasicrystals, and it
confirms recent experimental findings. The qualitative difference between our
results and previous calculations is partly explained by the small number of
$\bm{k}$ points that has usually been included in the evaluation of the
density of states of periodic approximants of quasicrystals. It is also shown
that both the density of states of a small approximant of the three-dimensional
Penrose tiling and the density of states of the ideal two-dimensional Penrose 
tiling do have spiky features, which also partly explains earlier predictions.
\end{abstract}

\noindent
PACS numbers:

\noindent
{\tt 71.23.Ft} - Quasicrystals

\noindent
{\tt 71.20.-b} - Electron density of states and bandstructure of 
crystalline solids

\section{Introduction}

Since the discovery of quasicrystals (QC's) \cite{She84}, it has been an 
important issue in condensed-matter physics to find out which properties 
distinguish this type of material, which has long-range order but no lattice 
periodicity, from crystals and amorphous materials.
Theoretically this problem was studied by means of so-called approximants. 
An approximant is a lattice periodic
crystal with a large unit cell with a structure 
close to that of a QC. For icosahedral QC's
approximants are obtained if in a higher
dimensional description of a QC \cite{Jan88}, the golden mean $\tau
= \frac{1}{2} (\sqrt{5} + 1)$ is approximated by a fraction. By approaching
$\tau$ in a systematic way, a series of approximants can be obtained with ever
larger unit cells and ever better structural convergence with respect to the
ideal QC.
By studying such a series of approximants and meanwhile
monitoring convergence with respect
to unit-cell size, results may be obtained that are valid for an ideal QC.
Due to computational limitations
most {\it first principles} calculations \cite{Fuj89,Fuj93a,Fuj93,Tra94} were 
restricted to so-called $1/1$ approximants,
where the fraction indicates the approximation made for $\tau$, with, 
depending on the particular QC, approximately 150 atoms
per unit cell. These calculations were done by means of the
linear muffin-tin orbital (LMTO) method.
They led to the surprising prediction that the 
electronic density of states (dos) of QC's
is spiky, a property that would be specific for QC's. It started in
1989, when Fujiwara \cite{Fuj89} predicted on the basis of a calculation for
1/1 AlMn that the dos of crystalline and quasicrystalline
AlMn consists of a set of spikes.
In 1993, he found that the electronic dos of 
$1/1$ AlMn \cite{Fuj93a,Fuj93} and $1/1$ AlCuLi \cite{Fuj93a}
has dense spiky peaks of width 10-20~meV or less, 
which in his view suggested that spikes are generic in QC's and their 
crystalline approximants.
In the same year, Hafner and Kraj\v{c}\'{\i} \cite{Haf93} conjectured 
on the basis of results for $1/1$ AlZnMg that
the dos of an ideal QC is spiky.
In 1994, De Laissardi\`ere and Fujiwara \cite{Tra94} found spikes of width
14 meV in $1/1$ ALCuFe. They stressed that the spikes are in the first place
a consequence of quasiperiodicity, accentuated by the presence of transition
metal atoms.
In 1995, Kraj\v{c}\'{\i} and co-workers \cite{Kra95} found a spiky dos 
for $1/1$ AlPdMn and $2/1$ AlPdMn.
The latter dos was obtained by folding 
the eigenvalues of the Hamiltonian at the $\Gamma$-point with a Gaussian of 
width 25~meV. Apart from being interesting, it was
indicated that spikes may also be relevant to electronic transport in
QC's \cite{May93,Fuj93,Tra94}.
Disappointingly spikes have 
not been observed directly in experiments, although a sufficiently high 
resolution was obtained by amongst others Stadnik and 
co-workers \cite{StL,Sta97}
by means of ultrahigh energy resolution photoemission spectroscopy (5~meV), 
Dadydov and co-workers \cite{Dad} in tunnelling experiments (1~meV) and 
Escudero and co-workers \cite{Esc} in point contact and tunnelling experiments
(0.35~meV). By convoluting the dos of $1/1$
AlCuFe \cite{Tra94} with a Gaussian of full width at half maximum of 31.6~meV,
Stadnik and co-workers \cite{Sta97} showed that even at a relatively low 
resolution, spiky features should according to the prediction be observed near
the Fermi level. In ref. \ref{Zij} we gave an overview of explanations that 
have been proposed for the discrepancy between theory and experiment, which
we will not repeat here. Instead, we will focus on two explanations. The
first was proposed by Hafner and Kraj\v{c}\'{\i} \cite{Haf}. They indicated
the possibility that spikes are specific for small periodic approximants of
QC's and do not persist in ideal QC's. The second explanation
is that spikes are an artefact of the calculation \cite{Sta97,Hae98}.
We have studied these explanations by calculating the dos
for the two-dimensional (2D) and the three-dimensional (3D)
Penrose tiling, which are 
quasicrystalline model systems. The 3D Penrose tiling has 
icosahedral symmetry and is the basis framework for many realistic models of 
icosahedral QC's \cite{Fuj89,Haf93,Els,Haf,Fuj93}.
It is important to realise that these models do not take into account
different kinds of electrons, nor do they represent 
realistic local atomic configurations
in QC's, but they are quasiperiodic, 
and as a consequence of their simplicity it is possible to calculate
properties of high approximants.
Whereas in most LMTO studies results extrapolated for the quasiperiodic limit 
are based on 
calculations for a $1/1$ approximant, we could calculate the dos of the
3D Penrose tiling with a resolution of 10~meV, a result valid in the
quasiperiodic limit.
In this letter, we show that spikes present in a small approximant of the 3D
Penrose tiling, do not survive in the ideal QC, in agreement with the original
suggestion of Hafner and Kraj\v{c}\'{\i} \cite{Haf}. This result falsifies the
prediction that all QC's have a spiky dos.

\section{Model}

For the 2D Penrose tiling, we have considered
the same model as in ref. \ref{Zij}.
The 3D Penrose tiling \cite{Soc} is made with the 
cut-and-project method \cite{Jan88}. We consider three types of 
approximants.
First of all, if in the internal space
$\tau$ is approximated by $F_{n+1} / F_n$ with
$F_i$ a Fibonacci number, then a 
so-called a periodic $F_{n+1} / F_n$ approximant is obtained.
If $F_{n+2}$ is even, there is a body centring \cite{Jan88}.
Second, if in the internal space an additional translation is made
along a vector ($1 / F_n$) ($\frac{1}{2}$, $\frac{1}{2}$, $\frac{1}{2}$), 
another
$F_{n+1} / F_n$ approximant is formed, which has a non-symmorphic cubic
space group, {\it e.g.}, Pa$\bar{3}$. We will refer to these 
approximants as cubic approximants. Finally, 
taking different approximations
for $\tau$ in different directions in the internal space, we consider 
$\tau_x = F_n / F_{n-1}$, $\tau_y = F_{n+1} / F_n$ and $\tau_z = F_{n+2} /
F_{n+1}$. This gives non-cubic ($F_n / F_{n-1}$, $F_{n+1} / F_n$, $F_{n+2} / 
F_{n+1}$) approximants, of which the lattice is body-centred
orthorhombic. An important parameter for each 
approximant is $N$, the number of vertices of the approximant tiling per unit 
cell.
We studied the so-called vertex model, which has one electronic basis state on
each vertex and nearest neighbour hopping via the edges of the
tiling. We chose the on-site energy to be zero. The electronic basis states 
can be thought of as {\it s}-electrons. The Hamiltonian is:
\begin{equation}
H = \sum_{nm} \left| n \right> T_{nm} \left< m \right| +
\sum_n \left| n \right> V_n \left< n \right|,
\end{equation}
with $V_n = 0$ and $T_{nm} = 1~{\rm eV}$ if the vertices $n$ and $m$ are 
connected by an edge of the tiling and zero otherwise.
This model has been studied before \cite{Mar,Kr2,Rie,Zi2}, but so far it was 
not possible to draw definitive conclusions about the nature of the spectrum.
Kraj\v{c}\'{\i} and Fujiwara \cite{Kr2} have shown that at $E = 0$ there are
infinitely many, strictly localised states.
An important symmetry of this model is the bipartite property \cite{Rie,Zij},
as a consequence of which the density of states is symmetric with respect to
$E = 0$ \cite{Mar,Kr2,Rie}. Also, for the non-body-centred cubic and for the
non-cubic approximants, the bipartiteness of the lattice can be used to reduce
the problem of finding eigenvalues of $H$ from an $N \times N$
problem to an $N/2 \times N/2$ problem \cite{Rie}.

\section{Method}

In order to determine the dos, we 
perform a bandstructure calculation.
Here $N_k$, the number of $\bm{k}$ points calculated in the asymmetric unit, is
important.
For the evaluation of the dos, we follow roughly the same scheme
as in ref. \ref{Zij}: First we calculate the exact average dos
of a linearly interpolated \cite{Jep} bandstructure of an approximant in
intervals of width 0.2~meV, where the averaging ensures uniform convergence
\cite{Win}. Then we convolute with a Gaussian of full width at half maximum
of typically 10~meV, which reduces the error in the dos by 
smearing away all features on a scale smaller than 10~meV including effects of 
the finite averaging interval of 0.2~meV, but which does
not smear away spikes of width 10-20~meV. This last
step is not general practice, but we will argue that it is useful.
Maximal errors for the dos of approximants were extrapolated by 
making a systematic comparison of the results for various $N_k$ and 
simultaneously assuming asymptotic behaviour: the error in the dos
at a certain energy is the integral 
between the true constant energy surface and the polyhedron approximating it
\cite{Blo}, which in three dimensions converges
as $N_k^{-2 / 3}$ \cite{Blo} for sufficiently large $N_k$. For an ideal 
tiling, the maximal error was extrapolated assuming that the asymptotic 
behaviour of the error is proportional to $1 / N$
in a given series of approximants \cite{Ent}, which holds provided that 
the error with respect to $N_k$ is negligible, which was checked explicitly.

\section{Results}

\begin{figure}[t]
\epsfig{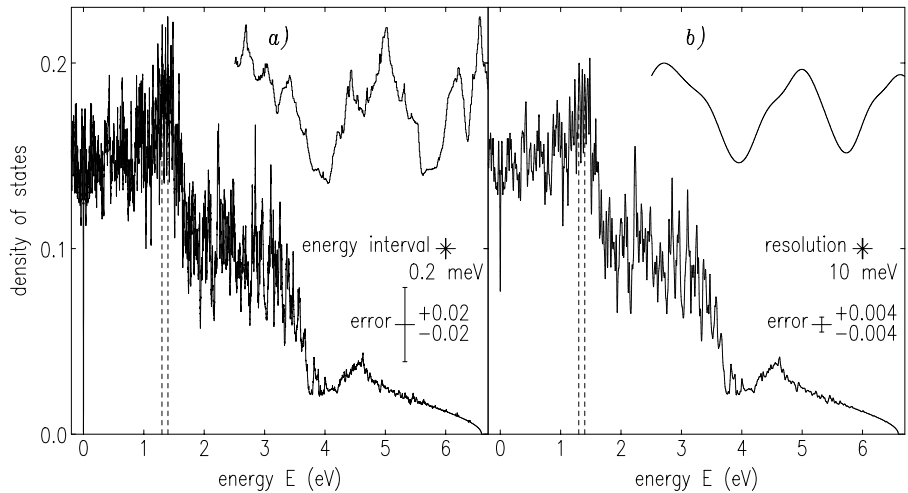}
\caption{The density of states of the cubic 3/2 approximant 
of the three-dimensional Penrose tiling {\it a)} before and {\it b)} after 
smoothening. The inset clarifies the energy range 1.3~eV to 1.4~eV.}
\label{f1}
\end{figure}

\begin{figure}[t]
\epsfig{file=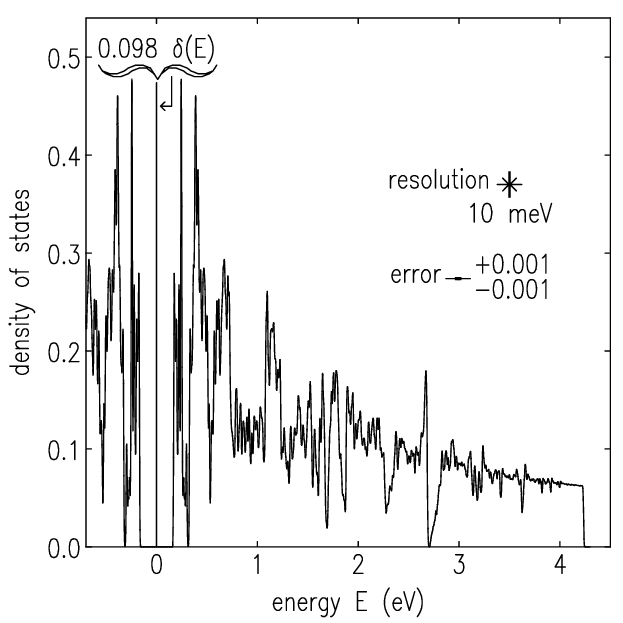,width=0.49\textwidth}
\hfill
\epsfig{file=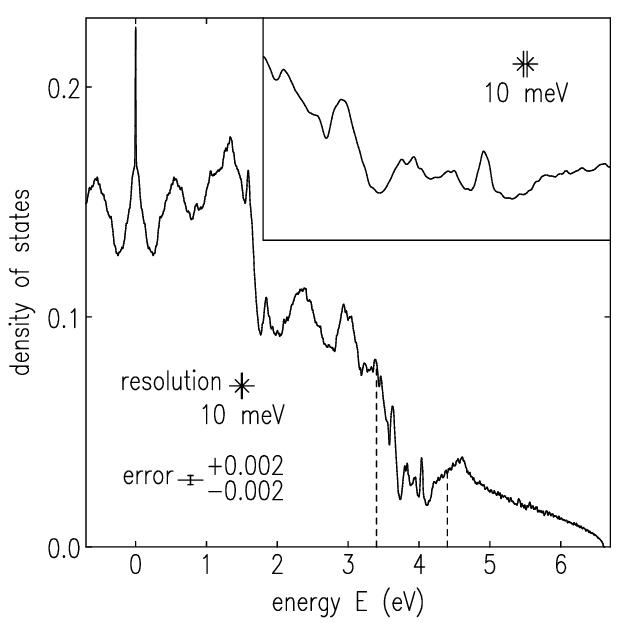,width=0.49\textwidth}
\caption{The density of states of the ideal two-dimensional Penrose tiling. The
delta peak at $E = 0$ is not smoothened.}\label{f2}
\caption{The density of states of the ideal three-dimensional Penrose tiling.
The inset shows the density of states in the energy interval 3.4~eV to 4.4~eV.}
\label{f3}
\end{figure}

\begin{figure}
\epsfig{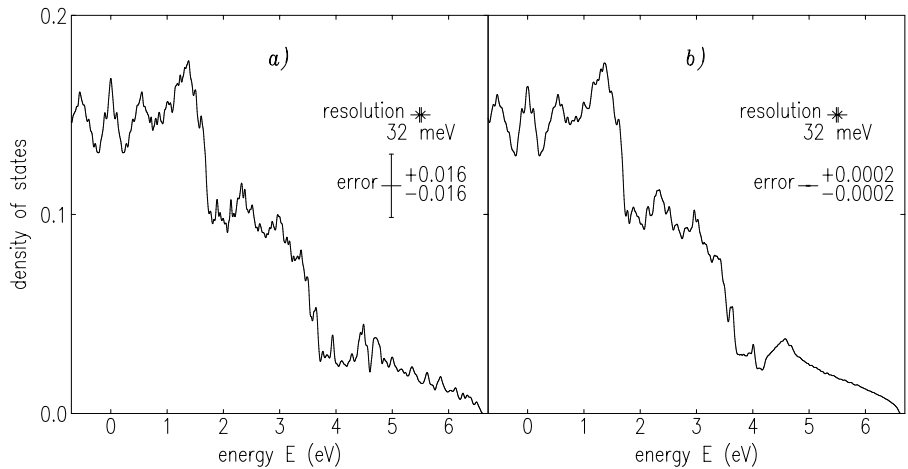}
\caption{The density of states of a (3/2, 5/3, 8/5) approximant of the
three-dimensional Penrose tiling with {\it a)} $N_k = 36$ and {\it b)}
$N_k = 16 388$.}
\label{f4}
\end{figure}

Figure \ref{f1}
shows the dos of the cubic 3/2 approximant of the 
3D Penrose tiling with $N = 576$ and $N_k = 1376$.
It has space group Pa$\bar{3}$. A grid of
so-called special $\bm{k}$ points \cite{Mon} was used. Figure 
1{\it a)} shows the result before convolution with a Gaussian: there are
many densely distributed spikes, but the error is so large that it is not
possible to say whether these spikes are real or an artefact connected to 
the small value of $N_k$. After smoothening, fig. 1{\it b)} is obtained.
The error is considerably smaller, so that conclusions can be drawn about the
spikes. Comparing the error and the amplitude of some of the spiky features,
we see that certainly some of these features are real.
So, the dos of small approximants of icosahedral 
QC's may have sharp features.
In fig. \ref{f2}, 
the dos of the ideal 2D Penrose
tiling is shown. It was calculated with $N=167 761$ and $N_k = 3$, where
$\bm{k}$ points were chosen that give real phase factors. The error indicates
convergence with respect to both $N$ and $N_k$. Our
conclusion of ref. \ref{Zij}, that the 2D Penrose tiling has a
spiky dos, is confirmed. 
In fig. \ref{f3},
the dos of the 3D Penrose tiling is shown. 
Actually, it is the dos of the non-cubic (8/5, 13/8, 21/13) 
approximant with
$N = 21 892$ and $N_k = 260$, where the $\bm{k}$ points lie on a regular grid 
which includes the $\Gamma$ point, but convergence with respect to $N$ was
checked. As far as our data can tell, this result does not depend on the
sequence of approximants that is considered. For example,
the 8/5 approximant with $N = 10 336$ and $N_k = 8196$
has a dos that deviates less than 0.008 from the graph of fig.
\ref{f3} for all energies except $E = 0$. Unfortunately, the
13/8 approximant with $N = 43 784$ is too large for us in order to be able 
to compute a sufficient number of $\bm{k}$ points.
Since the dos of the ideal
3D Penrose tiling in fig. \ref{f3} appears to be quite smooth,
we conclude that the spikes of fig. \ref{f1}{\it b)} do not survive in the 
quasiperiodic limit.
Figure \ref{f4}{\it a)} 
[t]
gives the dos of the non-cubic
(3/2, 5/3, 8/5) approximant of the 3D Penrose tiling with
$N = 1220$ and $N_k = 36$. The number of electronic basis states per unit
cell and $N_k$ were chosen as close as possible to those of the calculation
for icosahedral AlCuFe \cite{Tra94}, and we chose the same width for the
Gaussian as Stadnik and co-workers \cite{Sta97} did when smoothening it. 
Figure \ref{f4}{\it b)} shows the same dos
with $N_k = 16 388$. Comparing the two graphs, we see that taking into account
a too small number of $\bm{k}$ points may well lead to quite broad ($\sim$
32~meV) additional features in the dos. These features then
are an artefact. Because of the indicated error in fig. \ref{f4}{\it a)}
it is clear these features should not be taken seriously. The prediction for
AlFeCu of ref. \ref{Tra} does not indicate an error and from the published data
it is not clear whether the predicted spikes are an artefact or not. A too
small $N_k$ may well explain some of the spikes predicted in literature.
Expressly, we want to remark that it is in principle legitimate to try to keep
the number of $\bm{k}$ points that are calculated, small in order to save
limited computer resources, but one has to realise the price that is paid:
either the error is larger or the data should be presented with a worse
resolution.

\section{Conclusions}

We have carried out an analysis of the influence of the size of the 
approximant unit cell and the number of {\boldmath $k$} vectors in the
Brillouin zone on the finer features of the electronic dos in an icosahedral
tight-binding model.
We have shown that the dos of the 2D Penrose
tiling is spiky (fig. \ref{f2}), whereas the dos of the 
icosahedral 3D Penrose tiling is smooth (fig. \ref{f3}), at the given 
resolution of 10~meV.
We have also shown how a spiky dos would be predicted for our 3D model
when ({\it i}) only a small periodic approximant would have been studied 
(fig. \ref{f1}{\it b}) or ({\it ii}) a too small number of $\bm{k}$ points
would have been 
taken into account in the calculations, leading to large numerical errors
(fig. \ref{f4}).
Possibly the published calculations predictions that spikes are
generic for icosahedral QC's, have not sufficiently taken these two points
into account.
Our results further suggest that although experimentally spikes could not be
observed in QC's, they might be found in approximant phases.

\section*{Acknowledgments}

This work has been supported by the {\em Stichting voor Fundamenteel
Onderzoek der Materie} with financial support of the
{\em Nederlandse Organisatie voor Wetenschappelijk Onderzoek}.
We used free software from LAPACK \cite{Lap}. We thank Gilles de Wijs for
discussions.


\begin{thebibliography}{99}

\bibitem{She84} D. Shechtman, I. Blech, D. Gratias, and J. W. Cahn,
              Phys. Rev. Lett. {\bf 53}, 1951 (1984).
\bibitem{Jan88} T. Janssen, Phys. Rep. {\bf 168}, 55 (1988).
\bibitem{Fuj89} T. Fujiwara, Phys. Rev. B {\bf 40}, 942 (1989).
\bibitem{Fuj93a} T. Fujiwara, J. Non-Cryst. Solids {\bf 153-154}, 390 (1993).
\bibitem{Fuj93} T. Fujiwara, S. Yamamoto, and G. Trambly de Laissardi\`ere,
              Phys. Rev. Lett. {\bf 71}, 4166 (1993).
\bibitem{Tra94} G. Trambly de Laissardi\`ere and T. Fujiwara, Phys. Rev. B
              {\bf 50}, 5999 (1994). \label{Tra}
\bibitem{Haf93} J. Hafner and M. Kraj\v{c}\'\i, Phys. Rev. B {\bf 47},
              11 795 (1993). 
\bibitem{Kra95} M. Kraj\v{c}\'{\i}, M. Windisch, J. Hafner, G. Kresse, and
         M. Mihalkovi\v{c}, Phys. Rev. B {\bf 51}, 17 355 (1995).
\bibitem{May93} D. Mayou, C. Berger, F. Cyrot-Lackmann, T. Klein, and P. Lanco,
              Phys. Rev. Lett. {\bf 70}, 3915 (1993).
\bibitem{StL} Z. M. Stadnik, D. Purdie, M. Garnier, Y. Baer, A.-P. Tsai,
              A. Inoue, K. Edagawa, and S. Takeuchi, Phys. Rev. Lett. {\bf 77},
              1777 (1996).
\bibitem{Sta97} Z. M. Stadnik, D. Purdie, M. Garnier, Y. Baer, A.-P. Tsai,
              A. Inoue, K. Edagawa, S. Takeuchi, and K.H.J. Buschow,
              Phys. Rev. B {\bf 55}, 10 938 (1997). \label{Sta97}
\bibitem{Dad} D. N. Dadydov, D. Mayou, C. Berger, C. Gignoux, A. Neumann,
              A. G. M. Jansen, and P. Wyder, Phys. Rev. Lett. {\bf 77},
              3173 (1996).
\bibitem{Esc} R. Escudero, J. C. Lasjaunias, Y. Calvayrac, and M. Boudard,
              J. Phys.: Condens. Matter {\bf 11}, 383 (1999).
\bibitem{Zij} E. S. Zijlstra and T. Janssen, Phys. 
              Rev. B {\bf 61}, 3377 (2000). \label{Zij}
\bibitem{Haf} J. Hafner and M. Kraj\v{c}\'{\i}, 
              Phys. Rev. Lett. {\bf 68}, 2321 (1992).
\bibitem{Hae98} R. Haerle and P. Kramer, Phys. Rev. B {\bf 58}, 716 (1998).
\bibitem{Els} V. Elser and C. L. Henley,
              Phys. Rev. Lett. {\bf 55}, 2883 (1985).
\bibitem{Soc} J.E.S. Socolar and P.J. Steinhardt, Phys. Rev. B {\bf 34},
              617 (1986).
\bibitem{Mar} M.A. Marcus, Phys. Rev. B {\bf 34}, 5981 (1986).
\bibitem{Kr2} M. Kraj\v{c}\'{\i} and T. Fujiwara, Phys. Rev. B {\bf 38}, 
              12 903 (1988).
\bibitem{Rie} T. Rieth, Ph.D. thesis, Technical University Chemnitz-Zwickau,
              1996; T. Rieth and M. Schreiber, J. Phys.: Condens. Matter 
              {\bf 10}, 783 (1998).
\bibitem{Zi2} E. S. Zijlstra and T. Janssen,
              in Proceedings of the 7th International Conference on 
              Quasicrystals, Elsevier, 1999, edited by F. G\"ahler,
              P. Kramer, H.-R. Trebin, and K. Urban (in press).
\bibitem{Jep} O. Jepsen and O.K. Andersen, Solid State Commun. {\bf 9}, 1763
              (1971).
\bibitem{Win} R. Winkler, J. Phys.: Condens. Matter {\bf 5}, 2321 (1993),
              and references therein.
\bibitem{Blo} P. Bl\"ochl, O. Jepsen, and O. K. Andersen, Phys. Rev. B 
              {\bf 49}, 16 223 (1994).
\bibitem{Ent} O. Entin-Wohlman, M. Kl\'eman, and A. Pavlovitch, J. Phys. France
              {\bf 49}, 587 (1988). For the two-dimensional Penrose tiling, 
              the density of mismatches is shown to be proportional to $1 / N$.
              We assume the same behaviour for the three-dimensional Penrose 
              tiling.
\bibitem{Mon} H.J. Monkhorst, J.D. Pack, Phys. Rev. B {\bf 13}, 5188 (1976).
\bibitem{Lap} E. Anderson, Z. Bai, C. Bischof, S. Blackford, J. Demmel,
              J. Dongarra, J. Du Croz, A. Greenbaum, S. Hammarling,
              A. McKenney, and D. Sorensen, {\it LAPACK Users' Guide}
              (Society for Industrial and Applied Mathematics, Philadelphia,
              1999), 3rd edition.

\end{thebibliography}
\end{document}